\documentclass[aps,prd,twocolumn,groupedaddress, showkeys, showpacs]{revtex4}
\usepackage{graphicx,subfigure,enumerate,pstricks,latexsym}
\usepackage[english]{babel}

\providecommand{\dif}{\mathrm{d}} 
\def\beq{\begin{equation}}
\def\eeq{\end{equation}}
\def\bea{\begin{eqnarray}}
\def\eea{\end{eqnarray}}

\def\d{\dif}
\def\lightarrow{\rightarrow}

\def\SS{\Sigma}
\def\cosm{\lambda}
\def\parm{k}

\def\xx{\tilde{x}}
\def\yy{\tilde{y}}
\def\JJ{\tilde{J}}
\def\EE{\tilde{E}}
\def\rr{\tilde{r}}

\def\p{P}
\def\x{X}

\def\af{\zeta}
\def\tim{T}

\def\Veff{V_{\rm eff}}

\newcommand{\Schw}{Schwarzschild}

\graphicspath{{}}

\begin{document}

\title{Acceleration of string loops in the Schwarzschild-de Sitter geometry}
\author{Z. Stuchl\'{\i}k}
\author{M. Kolo\v{s}}
\affiliation{Institute of Physics, Faculty of Philosophy \& Science, Silesian University in Opava, Bezru\v{c}ovo n\'{a}m.13, CZ-74601 Opava, CzechRepublic}

\begin{abstract}
We study acceleration of current-carrying string loops governed by presence of an outer tension barrier and an inner angular momentum barrier in the field of Schwarzschild-de Sitter black holes. We restrict attention to the axisymmetric motion of string loops with energy high enough, when the string loop can overcome the gravitational attraction and escape to infinity. We demonstrate that string loops can be scattered near the black hole horizon and the energy of string oscillations can be efficiently converted to the energy of their linear motion. Such a transmutation effect can potentially represent acceleration of jets in active galactic nuclei and microquasars. We give the conditions limiting energy available for conversion onto the jet-like motion. Surpricingly, we are able to show that string loops starting from rest can be accelerated up to velocities $v \sim c$ even in the field of Schwarzschild black holes, if their angular momentum parameter is low enough. Such loops could serve as an explanation of highly relativistic jets observed in some quasars and active galactic nuclei. The cosmic repulsion becomes important behind the so called static radius where it accelerates the linear motion of the string loops up to velocity $v=c$ that is reached at the cosmic horizon of the Schwarzschild-de Sitter spacetimes independently of the angular momentum parameter of the strings. 
\end{abstract}

 
  
\maketitle


\section{Introduction}\label{intro}

Relativistic current carrying strings moving axisymmetrically along the axis of a Kerr black hole  \cite{Jac-Sot:2009:PHYSR4:} or a Schwarzschild---de Sitter (SdS) black hole \cite{Kol-Stu:2010:PHYSR4:} could in a simplified way represent plasma that exhibits associated string-like behavior via dynamics of the magnetic field lines in the plasma \cite{Sem-Ber:1990:ASS:,Chri-Hin:1999:PhRvD:,Sem-Dya-Pun:2004:Sci:} or due to thin isolated flux tubes of plasma that could be described by an one-dimensional string \cite{Spr:1981:AA:}. Tension of such a string loop prevents its expansion beyond some radius, while its worldsheet current introduces an angular momentum barrier preventing the loop from collapsing into the black hole. Such a configuration was also studied in \citep{Lar:1994:CLAQG:,Fro-Lar:1999:CLAQG:}. It has been proposed in \citep{Jac-Sot:2009:PHYSR4:} that this current configuration can be used as a model for jet formation. Here we shall test the possibility to converse motion of a string loop originally oscillating around a black hole in one direction to the perpendicular direction, modeling thus an accelerating jet. It is well known that due to the chaotic character of the  motion of string loops such a transformation of the energy from the oscillatory to the linear mode is possible \cite{Lar:1994:CLAQG:,Jac-Sot:2009:PHYSR4:,Kol-Stu:2010:PHYSR4:}; we estimate its efficiency and study the role of the cosmic repulsion. 

The relevance of the repulsive cosmological constant in the cosmological models was discussed in detail by \cite{Mis-Tho-Whe:1973:Gra:} or in \cite{Stu:2000:APS:,Boh:2004:GRG:}. Recent cosmological tests indicate presence of dark energy with properties close to those of nonzero (but very small) repulsive cosmological constant ($\Lambda > 0$) responsible for the observed present acceleration of the expansion of our universe \cite{Rie-etal:2004:ASTRJ2:}. More precisely, these cosmological tests indicate that the dark energy represents about $74.5\%$ of the energy content of the observable universe that is very close to the critical energy density $\rho_{\rm crit}$ corresponding to the almost flat universe predicted by the inflationary scenario \cite{Spe-etal:2007:ASTJS:}. Further, there are strong indications that the dark energy equation of state is very close to those corresponding to the vacuum energy, i.e., to the cosmological constant \cite{Cald-Kami:2009:NATURE:}. Therefore, it is  important to study the cosmological and astrophysical consequences of the effect of the observed cosmological constant implied by the cosmological tests to be $\Lambda \approx 1.3\times 10^{-56}{\rm cm}^{-2}$. It has been demonstrated that the repulsive cosmological constant is relevant in the framework of Einstein-Straus model of galaxy clusters immersed in expanding Universe \cite{Stu:1983:BULAI:,Stu:1984:BULAI:,Uza-Elli-Lar:2011:GRG:,Gre-Lak:2010:PHYSR4:} and even for matter orbiting supermassive black holes in Active Galactic Nuclei (AGN) since along with accretion discs also excretion discs have to be formed by the orbiting matter \cite{Stu-Sla-Hle:2000:ASTRA:,Stu-Sla:2004:PHYSR4:,Sla-Stu:2005:CLAQG:} being limited by the so called static radius behind which the cosmic repulsion prevails \cite{Stu-Hle:1999:PHYSR4:,Stu:2005:MPLA:}. Moreover, using the pseudo-Newtonian potential related to the SdS spacetimes \cite{Stu-Kov:2008:IJMPD:,Stu-Sla-Kov:2009:CLAQG:} it has been shown that the cosmic repulsion plays a strong role in motion of neighbourghing galaxies, as explicitly demonstrated for motion of Magellanic Clouds in the gravitational field of Milky Way \cite{Stu-Schee:2011:JCAP:}. Here we study acceleration of axisymmetric string loops in the field of SdS black holes that is interesting phenomenon by itself, and seems to be of crucial physical importance because of possibility to mimic acceleration of jets in AGN as discussed in \cite{Jac-Sot:2009:PHYSR4:}. 

\section{Current-carrying string loops}

We study a string loop threaded onto an axis of the SdS black hole spacetime chosen to be the $y$-axis - see Fig.~1 in \cite{Kol-Stu:2010:PHYSR4:}. The string loop can oscillate, changing its radius in the $x$-$z$ plane, while propagating in the $y$-direction. Assumed axial symmetry of the string loop allows to investigate only one point on the loop; one point path can represent whole string movement. Trajectory of the loop is then represented by a curve given in the 2D $x$-$y$ plane. The string loop tension and the worldsheet current corresponding to an angular momentum parameter form barriers governing its dynamics. These barriers are modified by the gravitational field of the SdS black hole characterized by the mass $M$ and the cosmic repulsion determined by the cosmological constant $\Lambda$. The SdS line element reads
\bea
 \d s^2 &=& -A(r) \d t^2 + A^{-1}(r) \d r^2 + r^2 (\d \theta^2 + \sin^2\theta \d \phi^2), \label{SfSymMetrika}
\eea
where the characteristic function takes the form
\bea
  A(r) &=& 1 - \frac{2 M}{r} - \frac{1}{3} \Lambda r^2.  
\eea
We use the geometric units with $c=G=1$ and the Schwarzschild coordinates. In order to properly describe the string loop motion, it is useful to use the Cartesian coordinates
\beq
 x = r \sin(\theta),\quad y = r \cos(\theta). \label{ccord}
\eeq
We summarize the equations of string motion in the Hamiltonian approach \cite{Lar:1993:CLAQG:}. 

\subsection{Equations of motion in Hamiltonian formulation} 

The string worldsheet is described by the spacetime coordinates $X^{\alpha}(\sigma^{a})$ (with $\alpha = 0,1,2,3$) given as functions of two worldsheet coordinates $\sigma^{a}$ (with $a = 0,1$). We adopt coordinates $(\tau, \sigma)$; $\tau$ denotes some affine parameter related to the proper time measured along the moving string, $\sigma$ reflects the axial symmetry of the oscillating string.

Dynamics of the string is described by the action related to the string tension $\mu > 0$ and a scalar field $\varphi$;
$ \varphi_{,a} = j_a $ determines current (angular momentum) of the string \cite{Jac-Sot:2009:PHYSR4:}. The assumption of axisymmetry implies the scalar field in linear form with constants $j_{\sigma}$ and $j_{\tau}$  
\beq
   \varphi = j_{\sigma}\sigma + j_{\tau}\tau.
\eeq

The worldsheet stress-energy tensor density $\SS^{ab}$ of the string can be expressed in the form \citep{Jac-Sot:2009:PHYSR4:}
\bea
 && \SS^{\tau\tau} = \frac{J^2}{g_{\phi\phi}} + \mu , \quad \SS^{\sigma\sigma} = \frac{J^2}{g_{\phi\phi}} - \mu, \nonumber \\
 && \quad \SS^{\sigma\tau} = \frac{-2 j_\tau j_\sigma }{g_{\phi\phi}}, \quad J^2 \equiv j_\sigma^2 + j_\tau^2.
\eea
Varying the action with respect to $X^{\alpha}$, we arrive to (second order differential) equations of motion \cite{Jac-Sot:2009:PHYSR4:,Kol-Stu:2010:PHYSR4:}
\beq
 (\SS^{ab} g_{\alpha\delta} X^\alpha_{,a})_{,b} - \frac{1}{2} \SS^{ab} g_{\alpha\beta,\delta} X^\alpha_{,a} X^\beta_{,b} = 0 . \label{EqOfMotionA}
\eeq

In general situations, integration of the equations of motion of string loops or open strings is a very complex task that has to be treated by numerical methods only, and has been discussed in a variety of papers \cite{Gar-Will:1987:PHYSR4:,Vil-She:1994:CSTD:,Lar:1994:CLAQG:,Fro-Lar:1999:CLAQG:,Pog-Vach:1999:PHYSR4:,Pag:1999:PHYSR4:,DeV-Fro:1999:CLAQG:,Sna-Fro-DeV:2002:CLAQG:,Fro-Fur:2001:PHYSR4:,Sna-Fro:2003:CLAQG:,Fro-Ste:2004:PHYSR4:}. The symmetry of the SdS spacetime and the assumption of the axisymmetric string oscillations enables a substantial simplification - the string motion can be treated using the Hamiltonian formalism. Following \citep{Lar:1993:CLAQG:}, we can introduce the Hamiltonian
\beq
 H = \frac{1}{2} g^{\alpha\beta} \p_\alpha \p_\beta + \frac{1}{2} \mu^2 r^2 \sin^2\,\theta + \mu J^2 + \frac{1}{2} \frac{(j_\tau^2 - j_\sigma^2)^2}{r^2 \sin^2\,\theta}, 
\eeq
where $\alpha, \beta$ correspond to coordinates $ t,r,\theta,\phi$. The spacetimes symmetries imply existence of two constants of motion
\beq
 \p_t = - E, \qquad \p_\phi = L = -2 j_\tau j_\sigma. \label{SAngMomentum}
\eeq
Then in spherically symmetric spacetimes the Hamiltonian can be expressed in the form
\beq
 H = \frac{1}{2} A(r) \p_r^2 + \frac{1}{2r^2} \p_\theta^2 - \frac{E^2}{2A(r)} 
     + \frac{\Veff(r,\theta)}{A(r)}, \label{HamHam}
\eeq
where an effective potential for the string motion has been introduced by the relation
\beq
\Veff(r,\theta) = \frac{1}{2} A(r) \left( \mu r \sin \theta  + \frac{J^2}{r \sin\theta} \right)^2.
\eeq
Notice that due to the symmetries of the spacetime and the oscillating string, the Hamiltonian is independent of the motion constant $L$, being dependent on $J^2$ only. 

Introducing an affine parameter of the string motion $\af$, the Hamilton–-Jacobi equations
\beq
 \frac{\d \x^\mu}{\d \af} = \frac{\partial H}{\partial \p_\mu}, \quad
 \frac{\d \p_\mu}{\d \af} = - \frac{\partial H}{\partial \x^\mu} \label{Ham_eq}
\eeq
applied to the Hamiltonian (\ref{HamHam}) imply equation of motion in the form
\bea
 \dot{r} = A \p_r, \quad \dot{\p_r} &=& \frac{1}{A} \frac{\p_\theta^2}{r^4} 
            \left( A r - \frac{1}{2} \frac{\d A}{\d r}   r^2  \right) \nonumber \\
            && - \frac{\d A}{\d r} \p_r^2  - \frac{1}{A}\frac{\d \Veff}{\d r}, \label{EqOfMotionB1} \\
 \dot{\theta} = \frac{\p_\theta}{r^2},  \,\,\,\quad  \dot{\p_\theta} &=&  - \frac{1}{A}\frac{\d \Veff}{\d \theta}. \label{EqOfMotionB2}
\eea
where dot means derivative with respect to the affine parameter: $\dot{f} = \d f/\d \af$. \footnote{Systems of equations of the string motion in the form of the second order differential equations (\ref{EqOfMotionA}) and first order equations (\ref{EqOfMotionB1}-\ref{EqOfMotionB2}) are related by the transformation  
$
 \d \tau = {\SS^{\tau\tau}} \d \af.
$
}

\subsection{Effective potential}

The condition $H=0$ determining regions allowed for the string motion \cite{Lar:1993:CLAQG:} can be written in the form
\beq
 E^2 = (\dot{r})^{2} + A(r) r^2 (\dot{\theta}^{2} ) + 2 \Veff. \label{StringEnergy}
\eeq
The loci where the string loop has zero velocity ($\dot{r}=0, \dot{\theta}=0$) form boundary of the string motion. We can define the boundary energy function by the relation 
\beq
 E^2_{\rm b} = 2 \Veff.
\eeq
The string loop motion is confined to the region where \cite{Kol-Stu:2010:PHYSR4:}
\beq
 \Veff(r,\theta) \leq 0.
\eeq

In Cartesian coordinates we arrive to the relation
\beq
E_{\rm b}(x,y;J) = \sqrt{A(r)}\left( \frac{J^2}{x} + x \mu \right) = \sqrt{A(r)} f(x), \label{EqEbXY}
\eeq
where $r = r(x,y) = \sqrt{x^2+y^2}$. The function $A(r)$ reflects the spacetime properties, while $f(x)$ those of the string loop. The behavior of the boundary energy function of the string motion is given by the interplay of the functions $A(r)$ and $f(x)$. Assuming that the string loop will start its motion from rest, i.e., assuming $\dot{r}(0) = 0$ and $\dot{\theta}(0)=0$, the initial position of the string will be located at some point of the energy boundary function $E_{\rm b}(x,y)$ of its motion.

It is obvious from equations (\ref{EqEbXY}) that we can make the rescaling $E_{\rm b} \rightarrow E_{\rm b} / \mu $ and $J \rightarrow J / \sqrt{\mu} $, assuming $\mu > 0$. This choice of ``units'' will not affect the character of the string boundary energy function, and is equivalent to the assumption of the string tension $\mu=1$ in Eq (\ref{EqEbXY}) - see \cite{Kol-Stu:2010:PHYSR4:}. In the following, we shall use this simplification. Moreover, it is convenient to introduce dimensionless coordinates and motion constants $\xx=x/M, \yy=y/M, \rr=r/M, \JJ=J/M, \EE=E/M$. We can assume $\JJ>0$ due to the spherical symmetry of the spacetime, similarly to the case of the motion of test particles. The detailed discussion of the properties of the effective potential and the string loop motion can be found in \cite{Kol-Stu:2010:PHYSR4:}. Here we summarize some results relevant for discussion of the string loop acceleration.

\begin{figure}
\includegraphics[width=0.8\hsize]{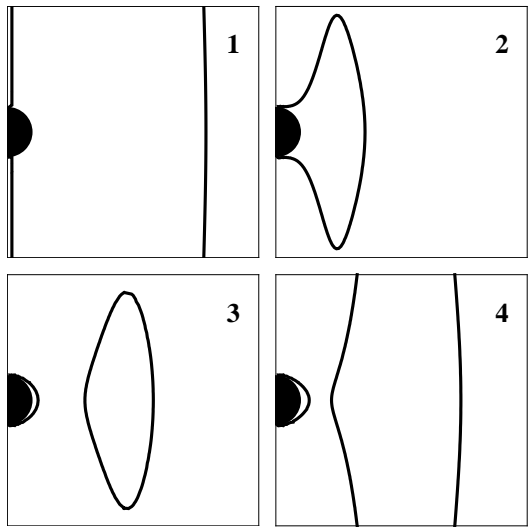}
\caption{\label{string_clas} Four different types of the behavior of the boundary energy function $E_{\rm b}(\xx,\yy;\JJ)$ in the Schwarzschild spacetimes; the fourth case has a subcase with the internal boundary enabling capturing of the string loop by the black hole, see \cite{Jac-Sot:2009:PHYSR4:, Kol-Stu:2010:PHYSR4:}.  }
\end{figure}
%
\subsection{String loops in the Schwarzschild geometry }
%

\begin{figure*}
\includegraphics[width=\hsize]{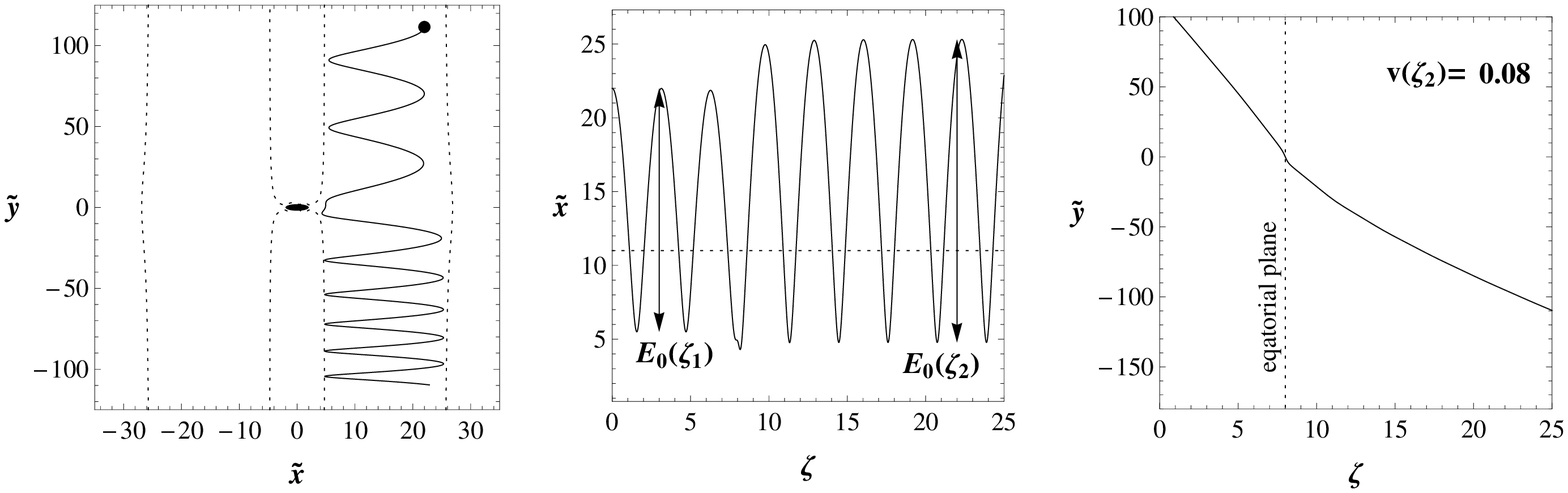}
\includegraphics[width=\hsize]{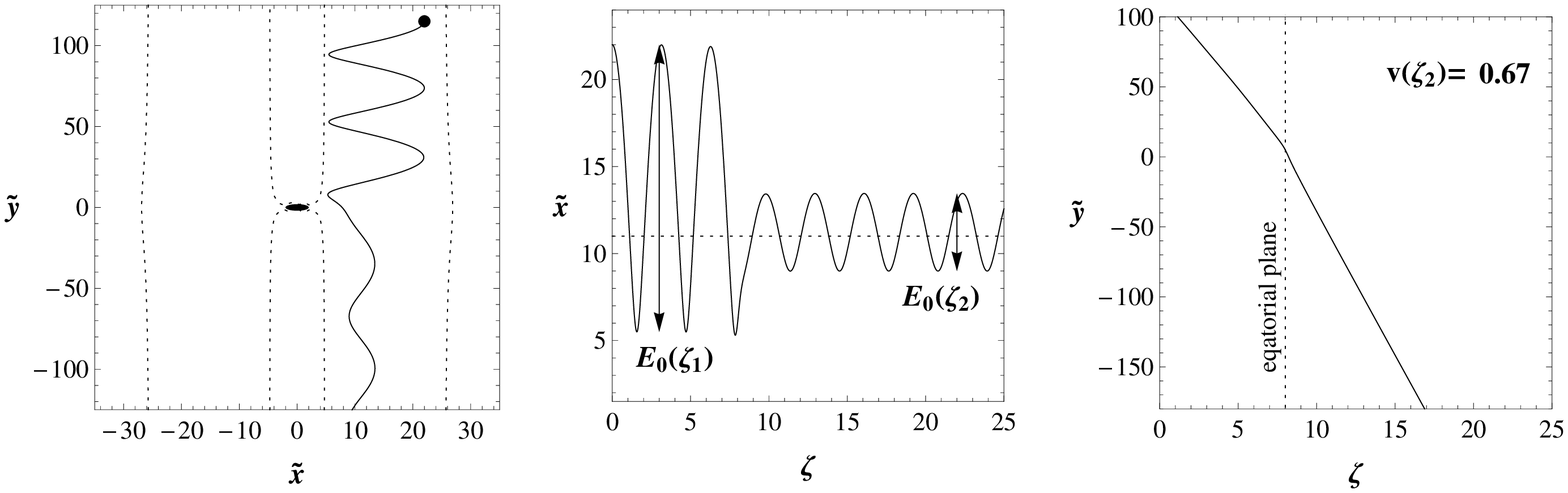}
\caption{String transmutation effect. Acceleration of the string motion in the $\yy$-direction (first row). Decceleration of the string motion (second row). Thick lines represent the string trajectory, while thin dotted lines in the first column correspond to the boundary energy function of the string motion $E_{\rm b}$. The string is assumed with angular momentum (current) parameter $\JJ = 11$ and starting at large distance from the black hole horizont, at $\xx=22$, and $\yy = 115$ (first row), $\yy = 111.5$ (second row). Near the starting point the spacetime is almost flat, so the string oscillates in the $\xx$-direction, while moving towards the black hole with initial speed in the $\yy$-direction $v \doteq 0.41$. For the affine factor $\af \sim 8$, the string approaches the region of strong gravity near the black hole, where the transmutation regime begins, and crosses the equatorial plane. The energy modes corresponding to the motion in the $\xx$- and $\yy$-direction are interchanging, and the string is chaotically scattered. In the first row acceleration of the string in the $y$-direction occurs with final velocity $v(\af_2) \doteq 0.67$, while in the second row decceleration occurs with final velocity $v(\af_2) \doteq 0.08$, where $\af_2 = 22.5$.
\label{transFig}}
\end{figure*}

\begin{figure*}
\subfigure[]{\includegraphics[width=5.5cm]{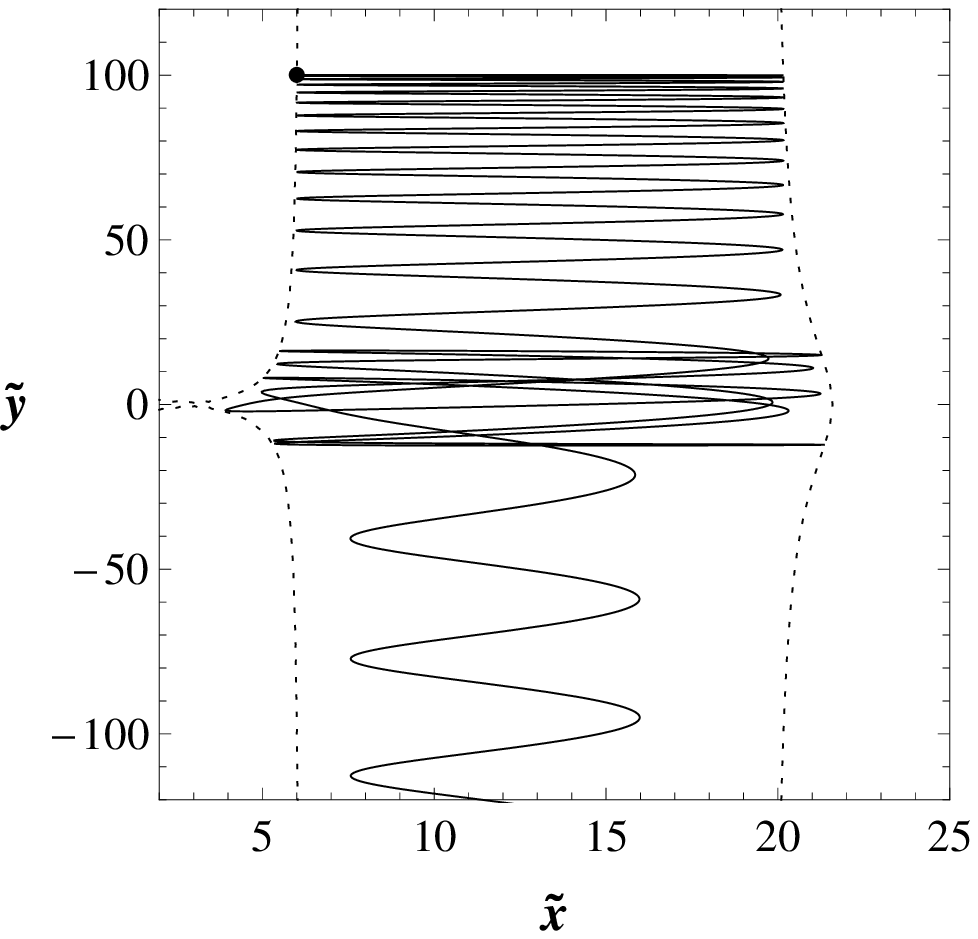}}
\subfigure[]{\includegraphics[width=5.5cm]{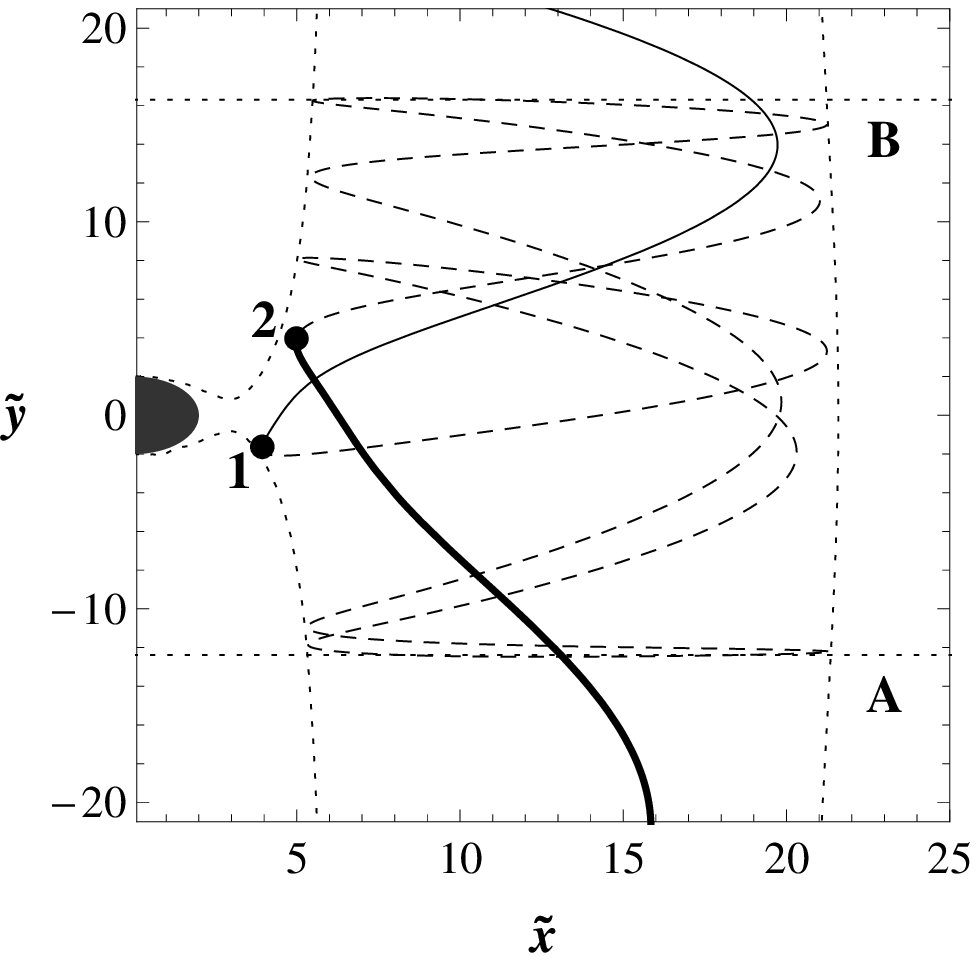}}
\subfigure[]{\includegraphics[width=5.5cm]{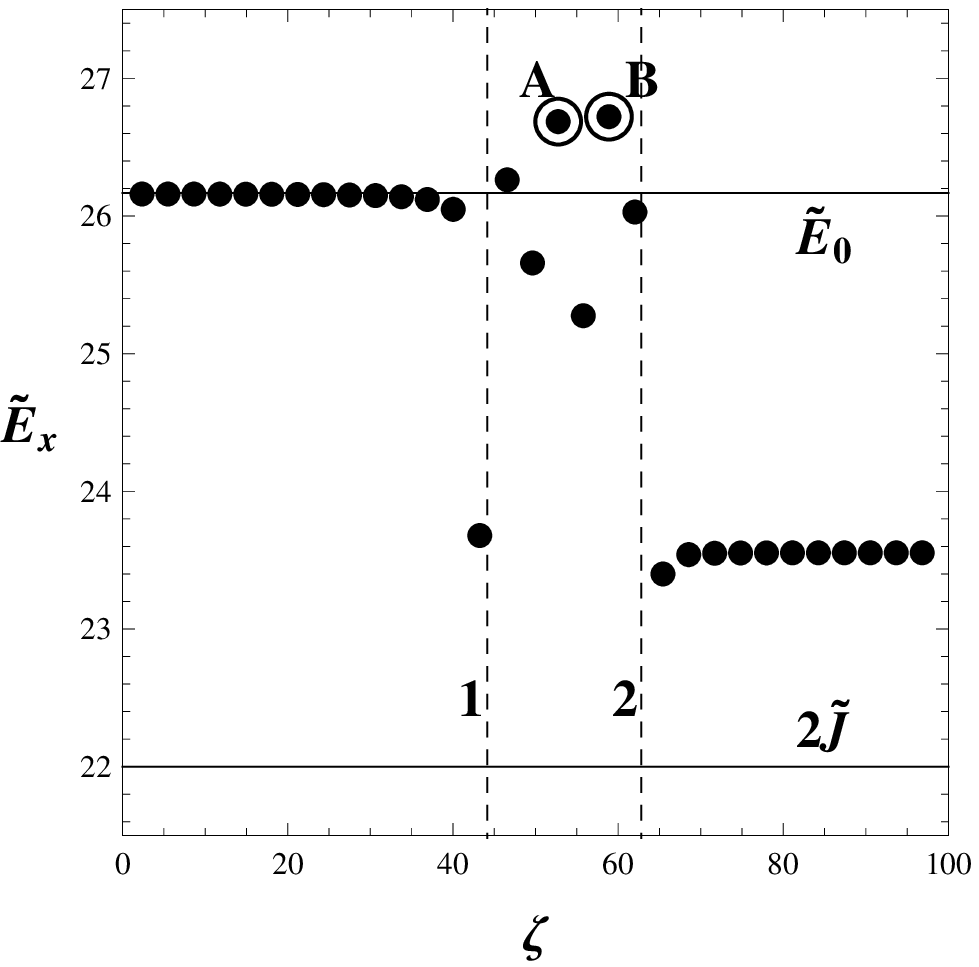}}
\vspace{-0.3cm}
\caption{ \label{stringFIG_3} 
{\bf Escape from the effective potential of black holes} String loop is starting at $\xx = 6, \yy = 100.1$ from the rest $\dot{\xx} = 0, \dot{\yy} = 0 $. String angular momentum and energy parameters are $ \JJ = 11,\EE = 25.9 $. At 
$\af \sim 40$ the string approaches the region near the black hole horizon (point (1)).  At points (A) and (B), the string cannot escape as the oscillatory energy is bigger than the boundary energy of the string at infinity. If the oscillatory energy is below this boundary energy, it can escape to infinity in the $\yy$-direction - see point (2) at time 
$\af \sim 65$. After the scatter, the string is moving in the $\yy$-direction with speed $v = 0.41$ ($\gamma = 1.099$). This is higher value of the transitional velocity than $v = 0.34$ presented in \cite{Jac-Sot:2009:PHYSR4:}. 
}
\end{figure*}

In the Schwarzschild spacetime the characteristic function of the line element (\ref{SfSymMetrika}) takes the form
\beq
            A(r) = 1 - \frac{2M}{r}.
\eeq
The Schwarzschild geometry introduces a characteristic length scale corresponding to the radius of the black hole horizon  $r_{\rm h} = 2M$. We restrict our considerations to the region above the black hole horizon, $r > r_{\rm h}$. In the Schwarzschild spacetime, there is $E_{\rm b} \lightarrow +\infty$ for $r \lightarrow \infty$, but $E_{\rm b} \lightarrow 0$ for $r \lightarrow 2M$. 

The extrema of the energy boundary function $\EE_{\rm b}$ are given by the relation \cite{Kol-Stu:2010:PHYSR4:}
\beq
      \JJ^2 = \JJ_{\rm E}^2(\xx) \equiv \frac{\xx^2(\xx-1)}{\xx-3}, \quad \yy = 0. 
\eeq
The extrema angular momentum function $\JJ_{\rm E}^2(\xx)$ diverges at $\xx = 3$ and at infinity. The minimal value of $\JJ_{\rm E}^2(\xx)$ is located at $\xx_{\rm min} \sim 4.303$ where the minimum takes the value of $\JJ_{\rm E(min)} \sim 7$; $\xx_{\rm min}$ determines marginally stable stationary position of string loops in the Schwarzschild spacetime - it is substantially lower in comparison with the innermost stable circular orbit (ISCO) of the free particle motion that is located at $\xx_{\rm ISCO} = 6$ \cite{Mis-Tho-Whe:1973:Gra:}.

The boundary energy function has two local extrema, maximum and minimum, located at $\xx > 2$, when 
\beq
    \JJ > \JJ_{\rm E(min)}.
\eeq
Then trapped states of oscillating strings can exist, corresponding to ``lakes'' of the effective potential (energy boundary function) from which the string loops may not escape to infinity neither may not be captured by the black hole. For appropriately chosen energy level of the oscillating strings, the region of trapping is determined by the energy boundary function $\EE_{\rm b}(\xx,\yy;\JJ)$. As demonstrated in \cite{Kol-Stu:2010:PHYSR4:}, we can distinguish four different types of the behavior of the boundary energy function and the character of the string loop motion in the \Schw{} spacetimes; we denote them by numbers 1 to 4 in Fig.~\ref{string_clas}. The first case corresponds to no inner and outer boundary; the string can be captured by the black hole or escape to infinity. In the second case, there is an outer boundary; the string loop cannot escape to infinity, it must be captured by the black hole. The third case corresponds to the situation when both inner and outer boundary exist and the string is trapped in some region forming a potential ``lake'' around the black hole. In the fourth case, the  string cannot fall into the black hole but it can escape to infinity (or be trapped). In the following, we are interested in situations corresponding to the cases 1 and 4 of the behavior of the energy boundary function enabling escape to infinity. Only these cases could represent an escaping string (jet).

\section{String loop transmutation and ejection speed}

The string transmutation effect occuring in strong gravitational field in vicinity of black holes means transmission between the energy of the oscillatory and linear translation motion of the string \cite{Lar:1994:CLAQG:,Jac-Sot:2009:PHYSR4:,Kol-Stu:2010:PHYSR4:}. The transmutation effect works in two ways as illustrated in Fig.~\ref{transFig}. In the first case amplitude of the string oscillations (in the $\xx$-direction) is lowered while the string motion is accelerated (in the $\yy$-direction). In this case some part the internal (oscillatory) energy of the string loop is transformed to the translation kinetic energy of the string (first row of Fig.~\ref{transFig}). The opposite case corresponds to amplitude amplification of the oscillations in the $\xx$-direction and decceleration of the linear motion in the $\yy$-direction; in this case the translation kinetic energy is partially converted to the internal oscillatory energy of the string (second row in Fig.~\ref{transFig}). We shall focus our attention to the case of accelerating string loops. We first study acceleration in the field of Schwarzschild black holes; in the next section the role of the repulsive cosmological constant is discussed and SdS spacetimes are considered. We shall discuss the situations when the energy of the translation motion enables escape of the string to infinity - concentraing especially on the possibility of string loops escaping with high velocities. Therefore, it is relevant to make decomposition of the energy of oscillating string loops in the flat universe (de Sitter universe when the cosmic repulsion is considered); such a decomposition has to be used in determining the behavior of strings moving far away from the black hole.

\subsection{Decomposition of the string energy in the flat spacetime}
Since the Schwarzschild metric is asymptotically flat, we first discuss the string loop motion in the flat spacetime. The energy of the string loop (\ref{StringEnergy}) in Cartesian coordinates reads
\beq
E^2 = \dot{y}^2 + \dot{x}^2 + \left( \frac{J^2}{x} + x \right)^2  =  E^2_{\mathrm y} + E^2_{\mathrm x}, \label{E2flat}
\eeq
where dot denotes derivative with respect to the affine parameter $\af$. We have introduced energy in the $x$- and $y$-directions by the relations
\beq
  E^2_{\mathrm y} = \dot{y}^2, \quad E^2_{\mathrm x} = \dot{x}^2 + \left( \frac{J^2}{x} + x \right)^2 = 
                                                          (x_{\rm i} + x_{\rm o})^2 = E^2_{0}. \label{restenergy}
\eeq
The energy in the $x$-direction is marked as $E_0$ and can be determined by the inner $x_{\rm i}$ and outer $x_{\rm o}$ radii limiting motion of the string loop
\beq
 x_{\rm o,i} = \frac{1}{2}\left( E_{0} \pm \sqrt{E_{0}^2 - 4 J^2} \right).
\eeq
The energy $E_0$ is minimal when the inner and the outer radii coincide - then 
\beq
 x_{\rm i} = x_{\rm o} = J 
\eeq
and the minimal energy is given by the minimal energy necesary for escaping of the string loop to infinity
\beq
 E_{\rm 0(min)} = 2 J. \label{E0min}
\eeq 
Clearly, $E_{\rm x}=E_{0}$ and $E_{\rm y}$ are constants of the string loop motion in the flat spacetime and no transmutation between these energy modes is possible there. On the other hand, in vicinity of black holes, the kinetic energy of the oscillating string can be transmitted into the kinetic energy of the translational linear motion. 

The energy in the $x$-direction $E_{0}$ can be interpreted as an internal energy of the oscillating string, consisting from the potential and kinetic parts; only in the limiting case of $x_{\rm i} = x_{\rm o}$, the internal energy has zero kinetic component. The string internal energy can in a well defined way represent the rest energy of the string moving in the $y$-direction in the flat spacetime.

\begin{figure*}
\includegraphics[width=\hsize]{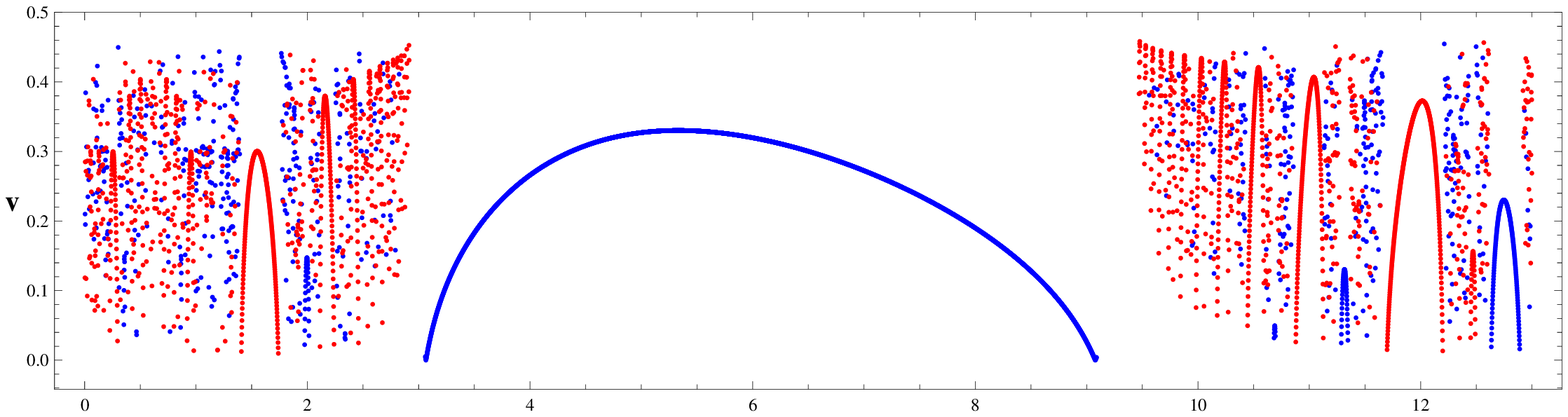}
\includegraphics[width=\hsize]{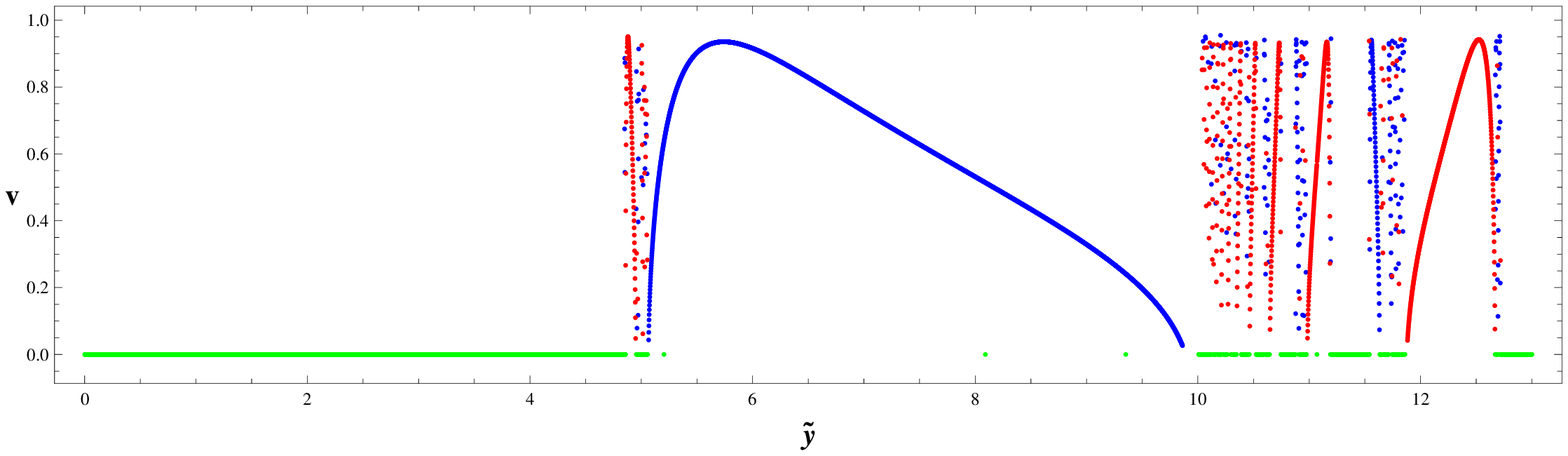}
\caption{Speed distribution of the string starting from the rest with $\JJ=11$ (top) and $\JJ=2$ (bottom) with the initial coordinate $\xx=20$, while the $\yy$-position (and total energy $\EE$) changes. The resulting velocities grow up to 
$v \sim 0.5$ (top) or $v \sim 1$ (bottom). Green colour represents trajectories captured by the black hole, blue colour corresponds to scattered escaped trajectories, $\yy\rightarrow-\infty$, while red colour denotes backscattered escaped trajectories, $\yy\rightarrow\infty$. \label{speedA} \label{speedB} 
}
\end{figure*}

\subsection{String ejection speed}

We assume an oscillating string loop with rest energy $E_{0}$ moving in the $y$-direction. The four-velocity norm condition  $U^\alpha U_\alpha = -1$ can be expressed in the form 
\beq
 g_{tt} \gamma^2 + g_{yy} u^2 = -1, \label{fourspeednorm}
\eeq
where the Lorentz factor $\gamma$ ($1 \leq \gamma < \infty $) and the $y$-component of the string four-velocity $u$ ($0 \leq u < \infty$) are given by the relations
\beq
 U^t = \frac{\d t}{\d \tim} = \gamma, \quad U^y = \frac{\d y}{\d \tim} = u. \label{speedty}
\eeq
$T$ is the proper time measured in the rest frame of the string loop.

In the flat spacetime with $g_{tt} = -1$ and $g_{yy} = 1$, Eqs. (\ref{fourspeednorm}) and (\ref{speedty}) imply the standard relations 
\beq
 \gamma^2 = u^2 + 1, \quad u = \gamma v, \quad \gamma^2 = \frac{1}{1-v^2},
\eeq
where the string loop coordinate velocity in the $y$-direction $v$ ($0 \leq v < 1$) related to the coordinate time $t$ reads
\beq
 v = \frac{\d y}{\d t}.
\eeq

If the string is not moving in the $y$-direction ($u=0$), the conserved component of its four-momentum (\ref{SAngMomentum}) in flat spacetime reads
\beq
 -E_0 = P_t = g_{tt} P^t = - \frac{\d t}{\d \af}, \label{nic1}
\eeq
where $E_0$ is the energy of a stationary string loop giving its internal rest energy, see (\ref{restenergy}). In the rest frame of a stationary string there is $\d t = \d \tim$ and we obtain from (\ref{nic1}) the relation 
\beq
 \d \tim = E_0 \d \af.
\eeq
Then (\ref{restenergy}) implies that for the transitional motion of a string loop there is \cite{Jac-Sot:2009:PHYSR4:} 
\beq
 E = \gamma E_0, \label{gamma}
\eeq 
where $E$ is the total energy of the string loop moving in the $y$-direction with the internal energy $E_{0}$. 
If we combine equations (\ref{E2flat}) and (\ref{gamma}), the $y$-component of the string four-momentum is given by
\beq
 P^y = \frac{\d y}{\d \af} = E_0 \frac{\d y}{\d \tim} = E_0 U^y.
\eeq
Clearly the energy $E_0$ plays the role of string rest energy. Using the Lorentz factor given by (\ref{gamma}), we can find the transitional velocity due to the relation
\beq
             v = \sqrt{1 - \frac{1}{\gamma^2}} .  \label{speed}
\eeq
In such a way the asymptotic energy (velocity) of the transitional motion of escaping string loops can be determined by the total energy $E$ and the internal (rest) energy $E_{0}$.

\begin{figure*}
\includegraphics[width=\hsize]{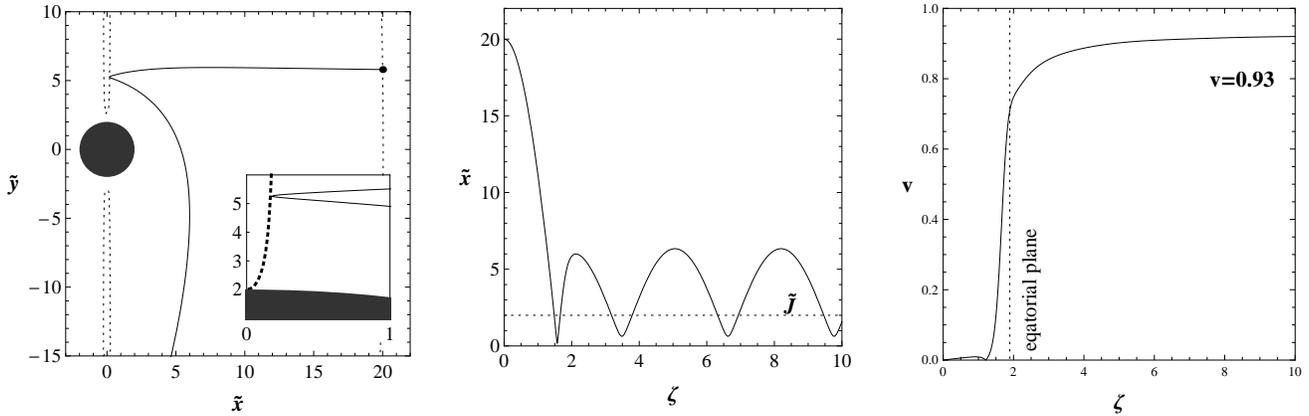}
\caption{Ultra-relativistic ejection of the string loop. The string is starting from rest relatively close to the equatorial plane, at $\xx_0 = 20, \yy_0 = 5.8$; with angular momentum parameter $\JJ = 2$ and energy $\EE \doteq 19.21$. We give the motion in the $\xx-\yy$ plane (a). In dependence on the affine parameter, we give evolution of the $\xx$-coordinate (b), and the velocity in the $\yy$-direction (c).
\label{ultra}}
\end{figure*}

\subsection{Transmutation effect in the Schwarzschild spacetime}

For a string with given angular momentum parameter $\JJ$ and total energy $\EE$, the maximal Lorentz factor (\ref{gamma}) and maximal transitional speed can be obtained, if the final energy in the $x$-direction (its value at infinity) is minimal, i.e., if $\EE_{\rm 0(min)} = 2 \JJ$. In order to obtain an acceleration of the string loop in the $\yy$-direction the string must past region near the black hole horizon, where string transmutation effect $\EE_{\rm x} \leftrightarrow \EE_{\rm y}$ occurs. In the Schwarzschild spacetime, $A(\rr) \neq 1$, we can express the string loop energy (\ref{StringEnergy}) in the Cartesian coordinates in the form
\bea
\EE^2 &=& A(\rr) \left( g_{xx}\dot{\xx}^2 + 2 g_{xy}\dot{\xx}\dot{\yy} + g_{yy}\dot{\yy}^2 \right) \nonumber \\
&&+ A(\rr) \, \xx^2 \left(\frac{\JJ^2}{\xx^2} - 1 \right)^2 ,  \label{Enedec1}
\eea
where the metric coefficients of the Schwarzschild spacetime in the $\xx$ and $\yy$-coordinates are given by
\bea
 && g_{xx} = \frac{\xx^2 + A \yy^2}{A (\xx^2+\yy^2)}, \quad g_{yy} = \frac{\yy^2 + A \xx^2}{A (\xx^2+\yy^2)}, \nonumber \\
 && g_{xy} = \xx \yy \frac{1 - A}{A (\xx^2+\yy^2)}.    \label{Enedec2}
\eea
The term $ g_{xy}\dot{\xx}\dot{\yy} $ is responsible for interchange of energy between the $\EE_{\rm x}$ and $\EE_{\rm y}$ energy modes - the string transmutation effect. The metric coefficient $g_{xy}$ is significant only in the neighborhood of the black hole, therefore, the effect of string transmutation can occur only in this region. 

All energy of the $\EE_{\rm y}$ energy mode can be transmitted to the $\EE_{\rm x}$ energy mode - oscillations of the string loop in the $\xx$-direction and the internal energy of the string will grow up maximally in such a situation, while the string will stop moving in the $\yy$-direction. On the other hand, all energy of the $\EE_{\rm x}$ mode cannot be transmitted to the $\EE_y$ energy mode - there remains inconvertible internal energy of the string, 
$\EE_{\rm 0(min)}=2\JJ$, being the potential energy hidden in the $\EE_{\rm x}$ energy mode.
 
The string transmutation in close vicinity of the black hole horizon chaotically changes the rate of the string propagation in the $\yy$-direction and, complementary, the internal energy $\EE_{0}$ related to the amplitude of the string oscillationary motion. An example of acceleration (decceleration) in the $\yy$-direction can be found in Fig.~\ref{transFig}. However, in order to have a proper physical insight into the transmutation effect, it is useful to follow the changes of the internal energy $E_{0}$ during the transmutation effect. This is illustrated in Fig.~\ref{stringFIG_3} for an illustrative example of a  string starting far away from the black hole horizon ($\xx = 6, \yy = 100.1$) at the rest state ($\dot{\xx} = 0, \dot{\yy} = 0$). The string loop parameters, angular momentum and energy ($\JJ = 11,\EE = 25.9$), are chosen to correspond to the case of the energy boundary function $\EE_{b}(\xx,\yy;\JJ)$ of the type 4, with barriere enabling capturing of the string by the black hole. Near the starting point, the spacetime is almost flat, so the string oscillates in the $\xx$-direction, but due to the gravitational attraction of the black hole the string is forced to move in the $\yy$-direction towards the black hole. The motion of the string is represented by subfigures a) and b) of Fig.~\ref{stringFIG_3}, while evolution of the internal energy $\EE_{0}$ of the oscillating string in terms of the affine parameter $\af$ is represented by subfigure c) of Fig.~\ref{stringFIG_3}.  For affine parameter $\af \sim 40$, the string approaches the region near the black hole horizon, where chaotic regime begins, and crosses the equatorial plane - see point (1). Near the black hole, the energy is succesively transformed between the $\xx$- and $\yy$-modes and the string is chaotically scattered. String can escape this region, falling down to the black hole or escaping to infinity in the $\yy$- or $-\yy$-direction. At the places (A) and (B), the string cannot escape, since its internal energy is larger than energy of the string at infinity - the string cannot fit the energy boundary function $\EE_{b}(\xx,\yy;\JJ)$ and must go back to the deeper part of the gravitational potential of the black hole. If the internal energy is below this boundary, the string can escape to infinity in the $\yy$-direction, see point (2) at the affine parameter $\af \sim 65$. After the period of chaotic scattering, the transmutation process is finished and the string is moving in the $\yy$-direction with transitional speed $v = 0.41$ and Lorentz factor $\gamma = 1.099$). 

Astrophysically most interesting situation corresponds to a string loop initially oscillating in (or near) the equatorial plane when its oscillatory energy is transmitted to the perpendicular direction; such a transmutation effect can represent  ejection of a jet from vicinity of black hole horizon. We discuss this toy model of jets considering strings starting from rest at small initial values of $\yy \leq 15$. The strings can escape to infinity in the $\yy$-direction, if the energy boundary function is of the type 1 or 4; for the type 2 and 3 there is a limit for the motion in the $\yy$-direction - see Fig.~\ref{string_clas} and discussion in \cite{Kol-Stu:2010:PHYSR4:}. Notice that very close vicinity of the equatorial plane is available only for the potential barriere of the type 4 when $\JJ > 9$ (see Fig. 7 in \cite{Kol-Stu:2010:PHYSR4:}.); in the case of the energy boundary function of the type 1, there is no repulsive barriere at radii close to the horizon and the string must start at sufficiently large distance in order to escape to infinity. We are interested in maximum of the trasitional speed in the $\yy$-direction that string loops can achieve asymptotically through the transmutation effect.

We give the study of the outcome of the transmutation process for both the relevant cases of the behavior of the energy boundary function. For the potential barreire of the type 4 with capturing of the string being forbidden, that is constructed for the angular momentum parameter $\JJ=11$, the results are demonstrated in Fig.~\ref{speedA} (top). The largest velocities for the string ejection reported in \citep{Jac-Sot:2009:PHYSR4:} are $v = 0.39$. On the other hand, in our study of strings with $\JJ > 9$, we have found higher values of the final velocity, growing up to 
$v = 0.5$. The results presented in Fig.~\ref{speedA} (top) clearly demonstrate the chaotic nature of the string transmutation effect. Notice that the regular part of the results of the simulations (in the region $3<\yy<9$) gives maximal $v \sim 0.3$, or $v \sim 0.4$ for $\yy \sim 11$, in accord with results of \citep{Jac-Sot:2009:PHYSR4:}, while the chaotic region allows $v \sim 0.5$. Notice that relatively large velocities $v \sim 0.5$ can be obtained for string loops starting very close to the equatorial plane.

There is an important question, whether ultrarelativistic transitional velocities could be achieved, and under which conditions. We are able to show that this is possible for small string angular momentum parameters, 
$\JJ < 9$. Then we have type 1 of the energy boundary function and any string starting close to the equatorial plane will collapse to the black hole - see Fig.~1. This implies necessity to start the string motion in sufficiently large distance from the equatorial plane. The results of modeling chaotic string loop motion finishing at infinity for $\JJ < 9$ is demonstrated in Fig.~\ref{speedB} (bottom) for $J=2$. We can see that now even in the regular part of the scattering process ultrarelativistic transitional velocities $v \sim 1$ occur. However, the "safe" initial  distance of string loops starts at $\yy=5$; for smaller distances the string is necessarily captured by the black hole. The acceleration to ultrarelativistic transitional velocity is demonstrated in Fig.~\ref{ultra}, where we show also dependence of the velocity of the escaping string loop on the affine parameter; the dotted line represents $\xx_{\rm i} = \xx_{\rm o} = \JJ$, i.e., the unconvertible energy of the string $\EE_{\rm 0(min)} = 2 \JJ$.
Notice the huge change of the oscillation amplitude which is responsible for the final acceleration up to the velocity $v \doteq 0.93$ in the $\yy$-direction.  Such  highly relativistic ejection speed is reported only for the type 1 of the energy boundary function (see Fig.~\ref{string_clas}), because the string loop has to go very closely to the black hole horizont where the string transmutation effect is strong enough. For the energy boundary function of the type 4, the largest ejection velocities we reported, $v \sim 0.5$, are substantially smaller; there exists a boundary preventing the string to go close enough to the black hole. 

%
\section{Influence of the cosmological constant }
%
\begin{figure*}
\includegraphics[width=\hsize]{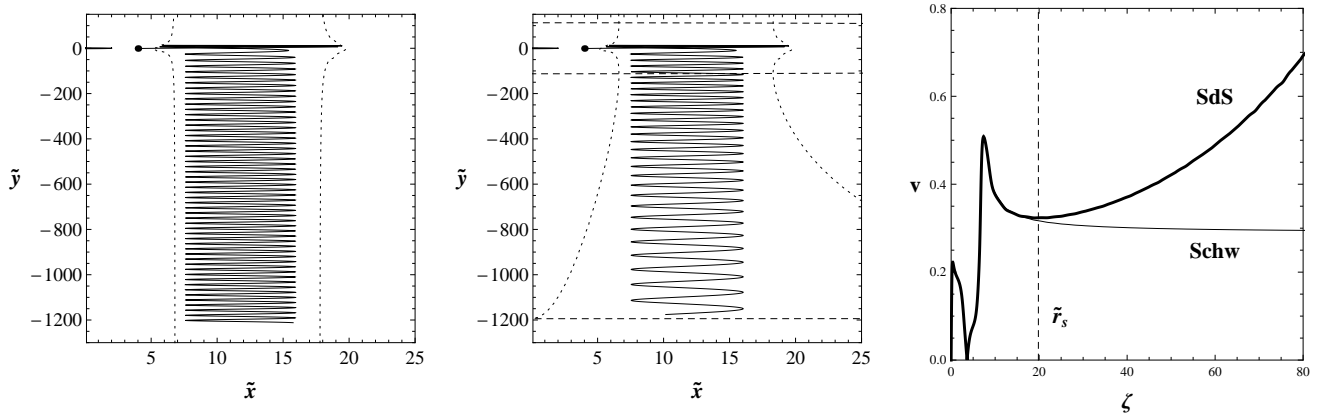}
\caption{Comparison of the string loop trajectories in the Schwarzschild (left) and the SdS spacetime (middle). The string parameters are $\JJ=11$ and $\EE \doteq 24.58$. The starting point is at rest at $\xx=4$ and $\yy=-1$. We can see how the string is stretched due to the cosmic repulsion in the $\yy$-direction by making comparison of the development of the transitional velocity in the Schwarzschild and the SdS spacetimes (right). In the SdS spacetime (with the cosmological parameter $\cosm = 7 \cdot 10^{-7}$), there is minimum of the string transitional speed at the static radius $\rr_{\rm s}$ and re-acceleration of the string loops occurs above the static radius. \label{SandSdS}}
\end{figure*}

The SdS spacetime is characterized by a dimensionless cosmological parameter 
\beq 
		\cosm = \frac{1}{3} \Lambda M^2.
\eeq 
In order to clearly demonstrate the role of the cosmological constant, we will use in the following the cosmological parameter $\cosm= 7 \cdot 10^{-7}$. Of course, this value is very large in comparison with values expected for realistic supermassive black holes in active galactic nuclei when even for the most extreme case of quasar $TON 618$ with mass of the central black hole estimated to be $M \sim 6.6 \times 10^{10} M_{\odot}$ there is $\cosm \sim 10^{-24}$ \cite{Ziol:2008:CJA:,Stu-Sla-Kov:2009:CLAQG:}, but it enables to demonstrate clearly the role of the cosmic repulsion.  

For $\cosm < 1/27$, there are the cosmological $\rr_{\rm c}$ and the black hole $\rr_{\rm h}$ horizons that are given by the relations \cite{Stu-Hle:1999:PHYSR4:,Stu-Sla-Hle:2000:ASTRA:}
\beq
 \rr_{\rm h} = \frac{2}{\sqrt{3 \cosm}} \cos\frac{\pi+\xi}{3}, \quad \rr_{\rm c} = \frac{2}{\sqrt{3 \cosm}} \cos\frac{\pi-\xi}{3}
\eeq
where
\beq
  \xi = \cos^{-1} \left( 3 \sqrt{3\cosm} \right).
\eeq
For $\cosm = 1/27$ the horizons coalesce at $\rr = 3$, while for $\cosm > 1/27$ the SdS spacetime describes a naked singularity. For astrophysically realistic black hole spacetimes ($\cosm < 10^{-24}$), there is with high precision $r_h \sim 2M$ and $r_c \sim \sqrt{3/\Lambda}$. The photon circular orbit is located at $\rr_{\rm ph} = 3$, independently of the cosmological parameter \cite{Stu:1990:BULAI:}. 

A crucial role plays the so called static radius
\beq
     \rr_{\rm s} = \cosm^{-1/3}
\eeq
where the gravitational attraction of the black hole acting on a test particle is just balanced by the cosmic repulsion \cite{Stu-Hle:1999:PHYSR4:}.

In the SdS spacetime, the asymptotic behavior of the boundary energy function is determined by the presence of the black-hole horizon and the cosmological horizon - there is $E_{\rm b} \lightarrow 0$ for both $r \lightarrow r_{\rm h}, 
r_{\rm c}$. 

The local extrema of the energy boundary function in the $\xx$-direction are determined by the relation 
\beq
  \JJ^2 = \JJ_{\rm E}^2 \equiv \frac{\xx^2(\xx-1-2\cosm \xx^3)}{\xx-3},
\eeq
while in the $\yy$-direction, the local minimum of the energy boundary function is located at $\yy=0$ and its local maximum is located at $\rr_{\rm max}=\rr_{\rm s}$. For values of $\cosm < \cosm_{\rm trap} \sim 0.00497 $, there are two maxima of the boundary energy function enabling oscillations in the $\xx$-direction \cite{Kol-Stu:2010:PHYSR4:}. Assuming $\cosm < \cosm_{\rm trap}$, we find that in the SdS spacetimes the classification of the behavior of the boundary energy function is similar to the classification relevant in the Schwarzschild spacetimes (with the types 1-4), giving captured, trapped and escaped motion (see Fig.~\ref{string_clas}) 
\footnote{In the SdS spacetimes there is, for some combination of the string parameters $\EE,\JJ$, an extra ``pathological'' case of the string-loop motion - the string-radius is exponentially growing in the $\xx$-direction due to the cosmic repulsion \cite{Kol-Stu:2010:PHYSR4:}.}.

\begin{figure*}
\includegraphics[width=\hsize]{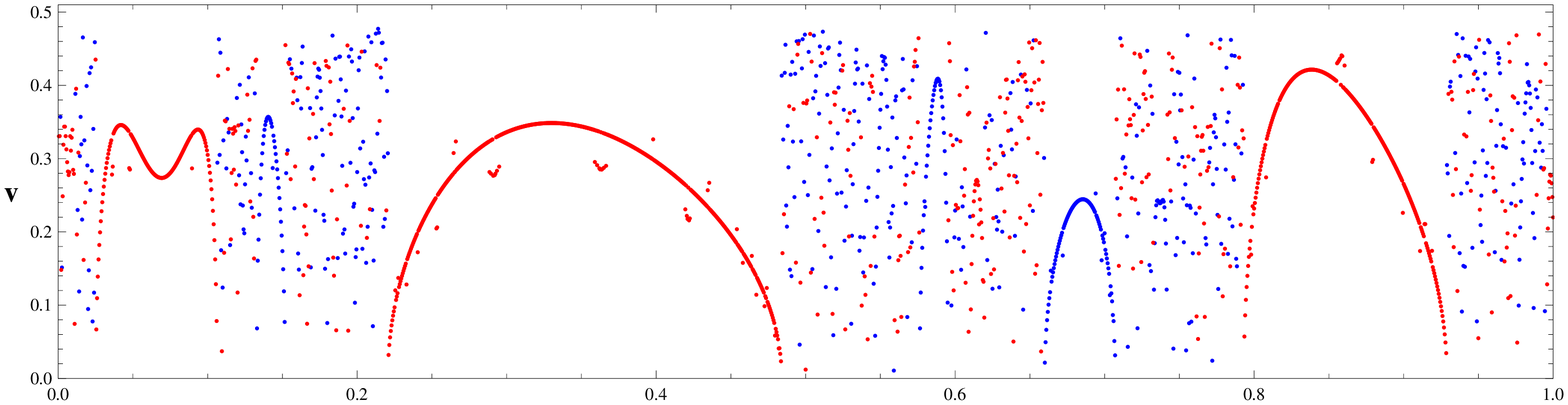}
\includegraphics[width=\hsize]{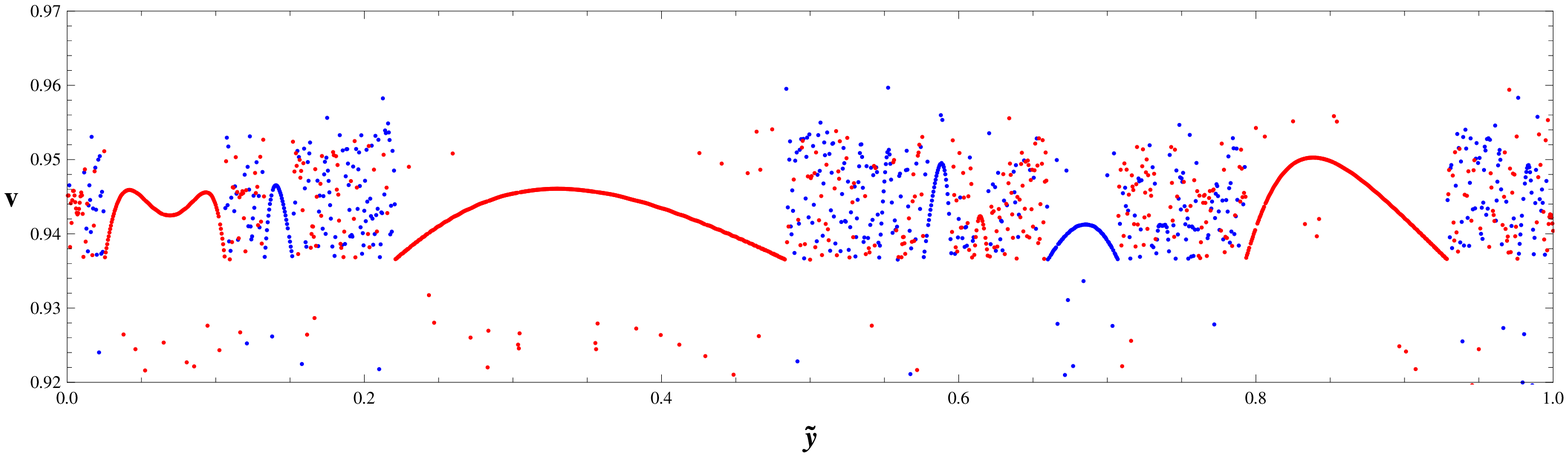}
\caption{Speed distribution of the string at the static radius $r_{\rm s}$ and near the cosmological horizon at 
$r = 0.8 r_{\rm c}$ in the SdS spacetime with $\cosm = 7 \cdot 10^{-7}$. The string starts from the rest with $\JJ=11$ and  initial coordinate $\xx=25$, while the initial $\yy$-position (and the total energy $\EE$) changes. There are no distinctive differences at static radius (upper figure) for the SdS and Schwarzschild spacetimes. On the contrary, near the cosmological horizon $r_c$ all string loops move with transitional speed near the speed of light, acceleration due to the repulsive cosmological constant is uniform and its influence on the "speed structure" is decisive. Maximal transitional  speed at $r = 0.8 r_{\rm c}$ is obtained to be $v\doteq0.96$, while its minimal value is $v\doteq0.92$. Blue colour represents escaped (scattered) trajectories, $\yy\rightarrow-\infty$, while red colour represents escaped (backscattered) trajectories, $\yy\rightarrow\infty$.
} \label{Statistic}
\end{figure*}

The maximum (``ridge'') of the boundary energy function $\EE_{\rm b}(\xx,\yy);\cosm,\JJ$) at $\rr_{\rm s}$ represents a boundary of the motion in the $\yy$-direction for string loops starting at $\rr < \rr_{\rm s}$. The lowest value of the boundary energy function at the static radius $ E_{\rm b}(\rr_{\rm s})$, giving the so called ``pass in the ridge'', is located at \cite{Kol-Stu:2010:PHYSR4:}
\beq
\xx_{\rm R}^2 = \JJ^2, \quad \yy^2 = \rr_{\rm s}^2 - \xx_{\rm R}^2.  
\eeq
The value of energy at the ``pass in the ridge'' is given by
\beq
 \EE_{\rm b(R)(min)} = 2 \JJ \sqrt{A(\rr_{\rm s})} = 2 \JJ \sqrt{1 - 3\cosm^{1/3}}.
\eeq
The string loop can escape to infinity if its energy overcomes the minimum energy on the ``ridge'', i.e., when $\EE > \EE_{\rm b(R)(min)}$.
Clearly, the static radius plays a fundamental role in the motion of string loops, similarly to the case of the motion of test particles. Because the string loops have to overcome the maximal value of the boundary energy $\EE_{\rm b}$ at $\rr_{\rm s}$, there will be minimum in the string loops speed in the $\yy$-direction, located just at the static radius. At 
$\rr > \rr_s$, cosmic repulsion accelerates the string loop in the $\yy$-direction.

\subsection{Decomposition of the string energy in the de Sitter spacetime}

The SdS spacetime is asymptotically de Sitter, not flat, therefore, we have to summarize properties of the string motion in the de Sitter spacetime and make decomposition of the string energy in a way similar to those applied for the flat spacetime. Since the center of the spherical symmetry of the de Sitter spacetime can be chosen at any point of the spacetime, similarly to the case of the flat spacetime, the amplitude of the oscillatory motion again remains constant \cite{Kol-Stu:2010:PHYSR4:}. We can demonstrate this fact formally by expressing the equations of the motion (\ref{EqOfMotionB1}-\ref{EqOfMotionB2}) in the $x$-$y$ plane for the de~Sitter spacetime:
\bea
  \ddot{x} - \frac{J^4}{x^3} + x &=& 2 \cosm x (J^2+x^2),  \label{eq05} \\
  \ddot{y} &=& 2 \cosm y (J^2+x^2). \label{eq06} 
\eea 
The equation for the oscillatory motion in the $x$-direction is independent of $y$ and its amplitude will remain constant during the motion in the $y$-direction. Of course, the amplitude is now influenced by the cosmological constant term as clear from the motion equation (\ref{eq05}) and character of the energy boundary function. We can see from the equations (\ref{eq05}-\ref{eq06}) that the string loop oscillates in the $x$-direction  between two turning points $x_{\rm i}, x_{\rm o}$, while moving and speeding up in the $y$-direction, due to the influence of the cosmological constant (for $\lambda=0$ acceleration in the $y$-direction vanishes). If $x_{\rm i} = x_{\rm o}$, the string is located at the minimum of the boundary energy function $E_{\rm b}$, being stationary in the $x$-direction. 

Introducing a new parameter  
\beq
     \parm = J \sqrt{\cosm} ,
\eeq
and assuming (quite naturally)
\beq
     \parm = \frac{J}{r_{\rm c}} << 1 ,
\eeq
we can obtain asymptotic formula for the location of the minimum of the energy boundary function $x_{\rm E(min)} \sim J $; for details see \cite{Kol-Stu:2010:PHYSR4:}. Then we can find analytical solution to the second equation of motion (\ref{eq06}) in the form
\beq
 y(\af) = y_0 \mathrm{Cosh}(2 \parm \af) + \frac{\dot{y}_0}{2 \parm} \mathrm{Sinh}(2 \parm \af) 
\eeq
where $y_0$ and $\dot{y}_0$ are initial values of integration of the motion equation.

In the de~Sitter spacetime, the string energy $E$ can be decomposed in the following way (\ref{StringEnergy})
\bea
 E^2 &=& \dot{x}^2(1-\cosm y^2) - 2 \cosm \, xy \, \dot{x} \dot{y} + \dot{y}^2(1-\cosm x^2)  \nonumber \\
 &&  + [1 - \cosm (x^2+y^2)]\left(\frac{J^2}{x} + x \right)^2.
\eea
The string rest (internal) energy $E_{0(dS)}$ takes the form 
\bea
 E_{\rm 0(dS)}^2 &=& (1-\cosm(x^2+y^2))\left( \frac{J^2}{x} + x \right)^2 \nonumber \\
 &=& (1-\cosm(x^2+y^2))( x_{\rm i} + x_{\rm o} )^2,  \label{E0deSitter}
\eea
and the Lorentz factor $\gamma$ (determining the string loop transitional velocity in the $y$-direction) is given by the formula
\beq
 \gamma^2 = \frac{E^2}{E^2_{\rm 0(dS)}} = \frac{E^2}{[1 - \cosm (x^2+y^2)](x_i + x_0)^2}. \label{gammaDS}
\eeq
The string loop velocity is then given again by Eq. (\ref{speed}).

When the string is not moving in the $\yy$-direction and only oscillates in the $x-z$ plane, the "rest" position of the string is equally defined at every point of the flat spacetime and the rest energy $E_{0}$ does not depend on the coordinates $x$ and $y$ - see Eq. (\ref{restenergy}). However, due the special character of the de~Sitter spacetime - expansion of the spacetime itself - the string rest energy $E_{\rm 0(dS)}$ given by Eq. (\ref{E0deSitter}) depends on the coordinates  $x$ and $y$, being modified by the metric factor $-g_{tt} = 1 - \lambda r^2$. We can see that string rest energy $E_{\rm 0(dS)}^2$ is positive only below the cosmological horizon; there is no "rest" state of the string above the horizon $r_{\rm c}$. Because the two turning points of the string loop oscillations $x_i$ and $x_o$ will remain constant, but the coefficient $[1 - \cosm (x^2+y^2)]$ is zero at the cosmological horizon $r_{\rm c}$, the Lorentz factor given by Eq. (\ref{gammaDS}) diverges at the cosmological horizon, and all string loops will pass the cosmological horizon with the speed of light. 

\subsection{Transmutation effect in the SdS spacetimes}

The transmutations of the string loop motion occur in the close vicinity of the black hole horizon, therefore, they are  determined by the phenomena demonstrated and studied in the case of the Schwarzschild geometry. The string loop energy decomposition in the SdS spacetimes is essentially identical with the decomposition formulae (\ref{Enedec1}) and (\ref{Enedec2}) used in the case of the Schwarzschild spacetime, since in the astrophysically relevant situations (with enormously small values of the cosmological parameter $\lambda$), the metric coefficients are essentially determined by the black hole mass parameter $M$ in vicinity of the black hole horizon. On the other hand, when the string crosses the static radius of the SdS spacetime, the cosmic repulsion becomes to be decisive for the string behavior; recall that at the static radius, the SdS spacetime is closest to the flat spacetime since their metric coefficients differ minimally - see \cite{Stu-Hle:1999:PHYSR4:}. At vicinity and behind the static radius, we have to use the expression for the energy decomposition derived for the de Sitter spacetime. We can expect substantial differences between the string loop behavior in the Schwarschild and SdS spacetimes behind the static radius.

We demonstrate the role of the repulsive cosmological constant in the transmutation effect by comparing the string loop motion in the Schwarzschild and SdS spacetimes for typical string angular momentum parameter ($\JJ=11$) corresponding to the case of string loops in states that do not enable them to collapse into the black hole, when their asymptotic speed in the flat spacetime is not strongly relativistic - see Fig.~\ref{SandSdS}. In such a case we can clearly see the strong role of the cosmic repulsion after the static radius of the SdS spacetime is crossed by the moving string loop. The speed of the moving string grows fastly at first stages of the expansion and then slowly decreases to the asymptotic value at the asymptotically flat Schwarzschild spacetime, while in the SdS spacetime it reaches a minimum at the static radius and then grows subtantially. We can see that acceleration of the string loop in the SdS spacetime is significant immediately after crossing the static radius, indicating potential observational effects in some AGN where the jets are reaching regions fairly exceeding extension of the associated galaxies.

A statistical study of the behavior of the string loops with $\JJ=11$ in the SdS spacetimes is presented in
Fig.~\ref{Statistic} that demonstrates velocities of the strings with different energy $\EE$ (starting at different initial positions due to changes of initial value of $y$) at the static radius where the SdS spacetimes is very close to the flat spacetime, and at $\rr = 0.8 \rr_c$ where the effects of the cosmic repulsion are very strong. We can convince ourselves that there are no relevant differences of the string loop velocities at the static radius obtained for the Schwarschild and SdS spacetime, but these are enornous at the radius $\rr = 0.8 \rr_c$ close to the cosmic horizon, where the velocities are highly relativistic, being in the interval of $0.92<v<0.96$. The distribution of the transitional velocities is more and more concentrated about the velocity of light with the string loop approaching the cosmic radius of the SdS spacetime.

\section{Conclusions}
 
We have studied possible acceleration of string loops in the field of Schwarzschild and SdS black holes due to the so called transmutation effect enabling transfer between energy of the oscillatory and transitional motion of the string loops. Such a process could serve as a toy model of acceleration of jets in quasars (AGN) or microquasars and is of astrophysical relevance. Our results can be summarized in the following way.

\begin{itemize}

\item
Conversion of energy between the oscillatory ($E_{\rm x}$) and transitional ($E_{\rm y}$) modes, i.e., the string transmutation effect, is sufficiently strong only in close vicinity of the black hole horizon. Transmision of energy between the oscillatory and transitional modes is of chaotic character and can occur multiply. The string can escape the black hole potential well only if amplitude of its oscillations can fit the energy boundary function $\EE_{\rm b}(\xx,\yy;\JJ)$ at infinity for the Schwarzschild spacetime, or the energy boundary function $\EE_{\rm b}(\xx,\yy;\lambda,\JJ)$ at the static radius for the SdS spacetimes.

\item
Our results clearly demonstrate that in the Schwarzschild spacetimes the largest ejection transitional velocities of the string loops starting from rest near the equatorial plane with large angular momentum parameter approach $v \sim 0.5$. Such a value is substantially larger than the value of $v \sim 0.39$ reported in \cite{Jac-Sot:2009:PHYSR4:} 

\item
Surprisingly, we have demonstrated that ejection speed $v \sim c$ can be obtained by the transmutation process in the field of Schwarzschild black holes, if the string loop with small angular momentum parameter starts from rest outside the equatorial plane, but close enough to it in order to enable the string loop to enter the deepest part of the potential well of the black holes where the transmutation effect is strongest.

\item
It is crucial that no rotation of the black hole is necessary for such extremely efficient acceleration of the jet-like motion due to the transmutation effect. On the other hand, near-extreme rotating black hole are usually considered if the Blandford---Znajek effect \cite{Bla-Zna:1977:MNRAS:} is assumed for acceleration of jet \cite{Pun:2001:BHGH:}.

\item
In the SdS spacetimes, the static radius $\rr_{\rm s}$ plays crucial role in the string acceleration by the transmutation process. The string loop has a minimal velocity at the static radius and is accelerated above it due to the effect of the repulsive cosmological constant. String loops cross the cosmological horizont $\rr_{\rm c}$ with the speed of light independently of their angular momentum parameter $J$. Considerable differences between Schwarzschild and SdS geometry can appear at relatively small distances above the static radius.

\end{itemize}

We can conclude that acceleration of string loops by the transmutation effect can serve as a toy model of acceleration of jets in quasars and AGN where ultra-relativistic speeds are observed. Of course, it can be applied also in the case of microquasars where the observed speeds are not ultrarelativistic. Really, the transmutation effect predicts acceleration of string loops to final velocities in whole interval of jet velocities observed in astrophysical objects. 

In the case of jets observed in AGN related to giant galaxies, the effect of the repulsive cosmological constant on the acceleration of string loops can be tested observationally. It is leading to re-acceleration of the string loops after crossing the static radius related to the central black hole of the AGN. The re-acceleration becomes sufficiently significant at relatively small distances from the static radius. The test is possible because extension of large galaxies related to AGN is comparable with the static radius related to their central supermassive black hole \cite{Stu-Sla-Hle:2000:ASTRA:,Stu:2005:MPLA:}, while the observed jets can substantially (several times) exceed extension of the associated  galaxies. 

The transmutation efect can be also astrophysically relevant while acting in the opposite direction, i.e., when the transitional energy of the string is converted to its oscillatory motion, since it can serve as a model of excitation of quasi-periodic oscillations observed in the innermost parts of the accretion discs orbiting black holes in microquasars or neutron stars in the low mass X-ray binary systems \cite{Tor-Abr-Klu-Stu:2005:ASTRA:,Tor:2005:ASTRA:,Tor:2009:ASTRA:}. We plan to present study of this phenomenon in a future work.

\section{Acknowledgments}
The present work was supported by the Czech Grant MSM~4781305903, and by the internal student grant of the Silesian University SGS/2/2010. 



\end{document}